\documentclass[11pt]{article}
\usepackage{amsfonts,latexsym}

\newcommand{\R}{\mathbb R}
\newcommand{\RR}{\mathbb R}

\newcommand{\calC}{\mathcal C}
\newcommand{\calN}{\mathcal N}
\newcommand{\calM}{\mathcal M}

\newcommand{\calL}{\mathcal L}
\newcommand{\calG}{\mathcal G}
\newcommand{\calF}{\mathcal F}

\newcommand{\del}{\partial}

\newcommand{\D}{{\mathcal D}}
\newcommand{\calD}{{\mathcal D}}

\newcommand{\ST}{\Sigma_{T}}

\newcommand{\g}{\gamma}

\newcommand{\gt}{\gamma_{T}}

\newcommand{\mut}{\mu_{T}}

\newcommand{\X}{{\bf X}}
\newcommand{\Y}{{\bf Y}}

\newcommand{\di}{\hbox{div}}
\newcommand{\dive}{{\mbox{div\,}}}
\newcommand{\tr}{{\mbox{tr\,}}}

\newtheorem{theorem}{Theorem}

\newtheorem{proposition}{Proposition}

\newtheorem{definition}{Definition}
\newtheorem{remark}{Remark}

\begin{document}

\title{On the topology of vacuum spacetimes}

\author{James Isenberg\thanks{Supported by the NSF under Grant 
PHY-0099373 and the American Institute of Mathematics}
\\ University of Oregon  \and
Rafe Mazzeo\thanks{Supported by the NSF under Grant DMS-9971975}
\\ Stanford University \and
Daniel Pollack
\\ University of Washington}

\date{September 6, 2002}

\maketitle

\begin{abstract}
We prove that there are no restrictions on the spatial topology
of asymptotically flat solutions of the vacuum Einstein equations in
$(n+1)$-dimensions. We do this by gluing a solution of the vacuum 
constraint equations on an arbitrary compact manifold $\Sigma^n$ to an 
asymptotically Euclidean solution of the constraints on $\R^n$. 
For any $\Sigma^n$ which does not admit a metric of positive scalar
curvature, this provides for the existence of asymptotically flat 
vacuum spacetimes with no maximal slices.
Our main theorem is a special case of a more general 
gluing construction 
for nondegenerate solutions of the vacuum constraint equations
which have some restrictions on the mean curvature, but for which
the mean curvature is not necessarily constant. 
This generalizes the construction \cite{IMP}, which is restricted to 
constant mean curvature data.
\end{abstract}

\section{Introduction}
\label{se:1}

A basic question in general relativity is whether there are any 
restrictions
on the topology of the spacetime manifold $M^{n+1}$ of a physically 
reasonable solution of Einstein's equations. If we restrict
attention to globally hyperbolic solutions, so that $M^{n+1}=\Sigma^n
\times \R$, and appeal to well-known results on the 
local well-posedness of the Einstein equations \cite{CB}, then
this question reduces to whether there are any restrictions on the 
topology of manifolds $\Sigma^n$ which carry physically realistic 
solutions of the Einstein constraint equations. The initial data
on $\Sigma$ is a pair of symmetric $2$-tensors $(\gamma, \Pi)$,
where $\gamma$ is a Riemannian metric and $\Pi$ represents
the second fundamental form in a Lorentzian development. 
The vacuum constraint equations are the compatibility conditions on these
initial data sets arising from the putative embedding in
a Ricci-flat Lorentz manifold. They take the form
\begin{eqnarray}
\dive \Pi - \nabla \tr \Pi & = & 0  \label{eqn1}\\
R - |\Pi|^2 + (\tr \Pi)^2 & = & 0 \label{eqn2}
\end{eqnarray}
All geometric quantities, norms and operators here are computed with respect
to $\gamma$, and in particular $R$ is the scalar curvature of this metric. 

We write $\tau = \tr \Pi$ and call this the mean curvature function
of the initial data set. Of particular interest and simplicity are the 
data sets with $\tau$ constant, and these are called constant mean 
curvature, or CMC. For $\Sigma$ compact, there are always CMC data sets 
on $\Sigma$ which satisfy the constraints :

\begin{proposition}
If $\Sigma^n$ is compact, then it admits solutions of the vacuum 
constraint equations (\ref{eqn1}) and (\ref{eqn2}) with constant mean 
curvature $\tau \neq 0$.
\label{cmc-compact}
\end{proposition}

In fact, any compact $\Sigma$ admits a metric $\gamma$ of constant 
scalar curvature $R = -n(n-1)$.  Setting $\Pi=\gamma$, 
then it is straightforward to check that $(\gamma,\gamma)$ solves
both (\ref{eqn1}) and (\ref{eqn2}) with $\tau=n$.  By rescaling we 
obtain solutions with $\tau$ an arbitrary positive constant. 

\medskip

Related to this result is the work of Witt \cite{Wi}. When $\Sigma$ is 
compact, any smooth function which is negative somewhere is the scalar 
curvature function for some metric on $\Sigma$ \cite{KW}, and he uses this 
to produce non-vacuum `dust' solutions to the constraints on $\Sigma$ 
with arbitrarily prescribed non-negative energy density.

Witt also addresses the question of whether there are topological 
restrictions for asymptotically Euclidean solutions of the constraints.
His results in this setting only hold 
for nonvacuum solutions, however. More specifically, he describes a 
procedure for gluing a solution of the constraints 
on an arbitrary manifold $\Sigma$ with nonvanishing 
energy density to a set of time symmetric initial data for a
Schwarzschild solution. This relies crucially on the nonvanishing of 
the energy density; hence his construction shows that there are 
asymptotically Euclidean solutions of the {\it non-vacuum} constraints 
on $\Sigma\setminus\{p\}$ for any compact manifold $\Sigma$,
but it says nothing about solutions of the vacuum constraints.
In fact, in his construction, the mass of the exterior Schwarzschild 
solution depends on the energy density of the interior, with the 
Schwarzschild mass equal to zero if the energy density vanishes. So if 
Witt's construction could be extended to the vacuum (non-flat) case, it 
would produce  nonflat solutions of the vacuum constraints which would be 
exactly Euclidean outside a compact set. This would violate the positive 
mass theorem \cite{SY2}. The recent work of Corvino \cite{Cor} shows 
that one can glue an exact exterior Schwarzschild metric 
(with non-zero mass) to compact 
subsets of fairly general time symmetric, asymptotically Euclidean 
initial data sets (with vanishing energy density). We emphasize, though,
that Corvino's construction begins with a pre-existing asymptotically 
Euclidean solution of the vacuum constraint equations. Hence, Corvino's 
work does not bear on the issue of topological restrictions for 
asymptotically Euclidean solutions.

Are there any restrictions on the topology of asymptotically Euclidean 
vacuum initial data sets? We shall prove that this is not the case.
\begin{theorem}
Let $\Sigma$ be any closed $n$-dimensional manifold, and $p\in \Sigma$ arbitrary.
Then $\Sigma\setminus\{p\}$ admits an asymptotically Euclidean initial data
set satisfying the vacuum constraint equations.
\label{mainthm}
\end{theorem}

We are {\it not} claiming that this solution is CMC. In fact, a CMC 
asymptotically Euclidean initial data set necessarily has
$\tau = 0$, so that (\ref{eqn2}) becomes 
\[
R=|\Pi|^2\geq 0.
\]
In addition, an asymptotically Euclidean metric on $\Sigma \setminus \{p\}$
with non-negative scalar curvature is conformally equivalent to (the
restriction of) a metric on $\Sigma$ with positive scalar curvature. 
This limits the possibilities for the topology of $\Sigma$
dramatically, cf.\ \cite{SY1}, \cite{GL1} and \cite{GL2}. For example, 
when $n=3$, this implies that $\Sigma$ is the connected sum of manifolds 
with finite fundamental group (i.e.\ the quotient of a homotopy sphere 
with positive scalar curvature by a finite group of isometries) and a 
finite number of copies of $S^2\times S^1$. The paper \cite{RS} surveys 
some of what is known in higher dimensions. 

Suppose we apply this construction when $\Sigma$ is a compact manifold 
which admits no metric of positive scalar curvature. We show in
\S 6 that, subject to an extra hypothesis which appears in 
the statement of Theorem~4, the maximal spacetime development 
of this data, which is asymptotically flat, admits no maximal slices. 
The existence of spacetimes with this property was heretofore unknown.

Theorem~1 is proved by an analytical gluing method closely
related to our earlier work \cite{IMP}. More specifically, we 
produce solutions on $\Sigma\setminus \{p\}$ by joining 
together a CMC solution on $\Sigma$ with a non-CMC solution 
on $\RR^n$. The main result of \cite{IMP} is that 
arbitrary nondegenerate (in a sense we explain below) solutions 
of the constraints on manifolds which are CMC and either compact,
asymptotically Euclidean or asymptotically hyperbolic
may be glued together. In particular, the method of \cite{IMP} shows
that if $\Sigma$ is any compact $3$-manifold, then 
$\Sigma \setminus \{p\}$ admits an asymptotically hyperbolic 
solution of the vacuum constraint equations. 
The gluing construction in the present paper closely follows
that earlier work, but the new feature here is the use
of non-CMC solutions on the asymptotically Euclidean summand. 
As we explain
in the next section, this complicates the analysis slightly
because the linearizations of (\ref{eqn1}), (\ref{eqn2})
uncouple when $\tau$ is constant, but not otherwise. 

In the next section we review the conformal method for solving
the vacuum constraints. In \S 3 we use the implicit function
theorem to find appropriate non-CMC asymptotically Euclidean 
solutions on $\RR^n$.
The gluing is done in two steps: a family of approximate solutions
is produced on $\Sigma {\#} \RR^n$, and then these are perturbed
using a contraction mapping argument to exact solutions.
These steps are reviewed in \S 4 and \S 5, respectively.
In \S6 we discuss the existence of asymptotically
flat vacuum spacetimes with no maximal slices.

A crucial idea in this analysis is the notion of nondegeneracy
of a solution, which concerns the surjectivity of the linearized
operator. We explain this concept in the next section. 

Theorem~1  is a special case of a more general gluing
theorem for non-CMC initial data sets. 

\begin{theorem} Let $(\Sigma_j,\gamma_j,\Pi_j)$, $j=1,2$,
be two initial data sets which solve the vacuum constraint
equations; these may be either compact, asymptotically
Euclidean or asymptotically hyperbolic. Suppose that both
solutions are nondegenerate with respect to the appropriate function
spaces (which contain functions weighted at infinity if
either factor is noncompact). If 
the mean curvature functions $\tau_j$ are both equal
to the same constant $\tau_0$ in a neighborhood of the
points $p_j \in \Sigma_j$, $j=1,2$, then the manifold 
$\Sigma_1 \# \Sigma_2$
obtained by forming a connected sum based at these two
points again carries a one parameter family of solutions of the 
vacuum constraint equations. Moreover, for large values of the
parameter, these solutions
are small perturbations of the original initial
data sets $(\gamma_j,\Pi_j)$ outside of small balls
around the points $p_j$.
\end{theorem}

This more general result is proved in almost exactly
the same way as the more specialized Theorem~1, and so we provide
details only in the special case. Note that in Theorem~2
one can also take the points $p_j\in \Sigma$, $j=1,2$ as lying in the 
same connected manifold.  In this case, rather than forming the connected 
sum, the theorem produces a family of solutions on the manifold obtained
from $\Sigma$ by adding a handle diffeomorphic to $S^{n-1}\times\R$.

The authors wish to thank the American Institute of
Mathematics, the National Science Foundation and the Stanford Mathematics 
Department for funding the extended Workshop on General Relativity in 
the Spring of 2002, during which this work was begun.

\section{The conformal method}
\label{se:3}
A very useful tool for the construction and enumeration of solutions to the 
vacuum constraint equations is the conformal method of
Lichnerowicz, Choquet-Bruhat and York, and many of the basic existence 
results for these equations, e.g.\ \cite{I1}, \cite{IM}, \cite{Ca1}, 
\cite{CCB}, \cite{COM}, \cite{CB2}, \cite{CIY}, \cite{AC}, rely on it. 
This method is most successful when dealing with CMC data because in
this case the equations decouple.
It can also handle non-CMC data, as shown in some of these references, 
and as our work shows here, although results to date suggest 
that one must place 
restrictions on the size of the gradient of the mean curvature. 

Solutions are constructed via the conformal method as follows.
One begins by fixing a background metric $\gamma$ (representing a 
given conformal structure), and a symmetric $(0,2)$ tensor
$\Pi$ which decomposes into trace-free and pure trace parts as 
$\mu + \frac{\tau}{n}\gamma$. The mean curvature function $\tau$
is specified through the second term on the right. (Note that one
often makes the additional demand that $\mu$ is transverse-traceless, 
i.e.\ also divergence-free. This is useful in parametrizing the set 
of solutions to the vacuum constraint equations but is misleading 
for our current purposes.)
We then modify this data by a conformal 
factor and a `gauging' term by setting
\begin{equation}
\tilde{\gamma} =  \phi^{\frac{4}{n-2}}\gamma, 
\qquad \widetilde{\Pi} = \phi^{-2}(\mu+\D W) + 
\frac{\tau}{n}\phi^{\frac{4}{n-2}}\gamma,
\label{eq:newdata}
\end{equation}
where $\phi$ and $W$ are a positive function and a vector field,
respectively. Note that the mean curvature is preserved, 
$\tau = \mbox{tr}_{\tilde{\gamma}}(\tilde{\Pi})$. 
The operator $\D$, which maps vector fields to trace-free symmetric 
$(0,2)$ tensors, is the conformal Killing operator, and is given 
in local coordinates by the formula
\[
\D X = \frac12 \calL_X \gamma - \frac{1}{n}(\di X) \gamma, \qquad
(\D X)_{jk} = \frac{1}{2}\left(X_{j;k}+X_{k;j}\right)-\frac{1}{n}\dive (X)\,
\gamma_{jk}.
\]
We have $\D X = 0$ if and only if $X$ is a conformal Killing field.
The formal adjoint of $\D$ on trace-free tensors is $\D^{*} = -\dive$ 
and the operator $L \equiv \D^{*}\circ\D$ is formally self-adjoint, 
nonnegative and elliptic. 

The modified data (\ref{eq:newdata}) satisfies the vacuum Einstein 
constraint equations (\ref{eqn1}) and (\ref{eqn2}) if and only if
\begin{eqnarray}
\Delta_{\gamma}\phi - \frac{n-2}{4(n-1)}R_\gamma\phi
+ \frac{n-2}{4(n-1)}\big|\mu+\D W\big|^2\phi^{\frac{-3n+2}{n-2}}-
\frac{n-2}{4n}\tau^2\phi^\frac{n+2}{n-2} &=& 0\label{lich} \\
LW -(\dive\mu-\frac{n-1}{n}\phi^{\frac{2n}{n-2}}\nabla\tau) &=& 0 \label{vecteqn}
\end{eqnarray}
The first of these is usually called the Lichnerowicz equation. 
We write this coupled system as $\calN(\phi,W,\tau) = 0$. 
The mean curvature $\tau$ is emphasized in this
notation; however, the 
dependence of $\calN$ on $\gamma$ and $\mu$ is suppressed.
The linearization $\calL$ of $\calN$ in the directions $(\phi,W)$ 
(but not $\tau$) is central to our construction. 

\begin{definition}
A solution to the constraint equations $\calN(\phi, W,\tau)=0$
is nondegenerate with respect to Banach spaces $\X$ and $\Y$
provided $\calL: \X \to \Y$ is an isomorphism.
\end{definition}

\begin{remark} It might seem more natural to require only
that $\calL$ be surjective. However, in the main cases of
interest, when $\Sigma$ is compact, asymptotically Euclidean
or asymptotically hyperbolic, these are equivalent (provided
we use spaces of functions which decay at infinity).
\end{remark}

Nondegeneracy conditions like this one are crucial to any
gluing construction. The main result of \cite{IMP} is that
any two nondegenerate solutions of the vacuum constraint equations 
with the same constant mean curvature $\tau$ can be glued.
For compact CMC solutions, nondegeneracy is equivalent to $\Pi\not\equiv 0$
together with the absence of conformal Killing fields.
On the other hand, asymptotically Euclidean or asymptotically 
hyperbolic CMC solutions are always nondegenerate (cf. \S7 of \cite{IMP}).
While the paper \cite{IMP} only treats the case $n=3$,  
the generalization to higher dimensions is not difficult; this is 
discussed in \cite{IMaxP}, which also considers the extension to 
various types of non-vacuum solutions.

Suppose $(\phi,W,\tau)$ solves  (\ref{lich}) and 
(\ref{vecteqn}) with background data 
$(\gamma,\Pi)$; then we can choose  the resulting solution 
$(\tilde{\gamma},\tilde{\Pi})$ of the constraints (\ref{eqn1}) and 
(\ref{eqn2}), defined by (\ref{eq:newdata}), as the new background 
data. For the moment, all objects with tildes are associated to this
new data. Let us determine the solution of 
$\tilde{\calN}(\cdot, \cdot)=0$ associated to 
this new data. Obviously the conformal factor $\tilde{\phi} = 1$,
but we need to find the new vector field. If we assume that 
our solution  $(\tilde{\gamma},\tilde{\Pi})$ is nondegenerate, then 
we may {\em uniquely} solve  $\tilde{L}\tilde{W} = 
(\widetilde{\dive}\mu-\tilde{\nabla} \tau$) (since $\tilde{\phi} = 1$) 
for $\tilde{W}$ in the appropriate summand of $\X$. 
Thus $\widetilde{\calN}(1,\tilde{W}) = 0$.  We drop 
the tildes henceforth.

%
%

\section{Non-CMC asymptotically Euclidean initial data on $\R^n$}
\label{se:2}
A metric $\gamma$ on $\R^n$ is said to be asymptotically Euclidean,
or  AE, if $\gamma$ decays to the Euclidean metric at some rate.
More precisely, we assume that there is a $\nu>0$ so that
in Euclidean coordinates $z$, $|\gamma_{ij}(z) - \delta_{ij}| 
\leq C|z|^{-\nu}$, along with appropriate decay of the derivatives,
as $z \to \infty$. To formulate this precisely, we make the
\begin{definition} 
The family of weighted H{\"o}lder spaces $\calC^{k,\alpha}_{-\nu}(\RR^n)$ 
is defined as follows:
\begin{itemize}
\item[]
\[
\calC^{0,\alpha}_0(\RR^n) =
\left\{u \in \calC^{0,\alpha}_{\mathrm{loc}}(\RR^n) \cap L^\infty(\RR^n):
\sup_{R\geq 1} \sup_{z \neq z' \atop R \leq |z|,|z'| \leq 2R}
\frac{|u(z)-u(z')|R^\alpha}{|z-z'|^\alpha}<\infty \right\}
\]
\item[]
\[
\calC^{k,\alpha}_0(\RR^n)=\{u: (1+|z|^2)^{|\beta|/2}\,\del_z^{\beta} u 
\in \calC^{0,\alpha}_0(\RR^n), \quad \mbox{for}\ \ |\beta|\leq k\},
\]
\item[]
\[
\calC^{k,\alpha}_{-\nu}(\RR^n)= \{u = (1+|z|^2)^{-\nu/2}v: 
v \in \calC^{k,\alpha}_0(\RR^n)\}.
\]
\end{itemize}
\end{definition}
Thus $u \in \calC^{k,\alpha}_{-\nu}(\RR^n)$ provided it is in the 
ordinary H\"older space $\calC^{k,\alpha}$ on any compact set
and also satisfies $|\del_z^\beta u| \leq C_\beta |z|^{-\nu-|\beta|}$
as $|z| \to \infty$, for $|\beta| \leq k$, with an appropriate decay condition 
for the $\alpha$ H\"older seminorm on the $k^{\mathrm{th}}$
derivatives of $u$. 

We define 
\[
{\mathcal M}^{k,\alpha}_{-\nu} = \{\mbox{metrics}\ \ \gamma\ 
\mbox{on}\ \RR^n: \gamma_{ij}-\delta_{ij} \in \calC^{k,\alpha}_{-\nu}\}.
\]
Note that if $\gamma \in {\mathcal M}^{k+2,\alpha}_{-\nu}$,
then its scalar curvature function $R_\gamma$ is in
$\calC^{k,\alpha}_{-\nu-2}$. Thus (\ref{lich}) and (\ref{vecteqn}) 
suggest that we should assume that
\[
W \in \calC^{k+1,\alpha}_{-\nu/2}(\RR^n,T\RR^n), \qquad \mbox{and}
\qquad \tau \in \calC^{k,\alpha}_{-\nu/2-1}(\RR^n).
\]

We now assume that $\mu \equiv 0$ and 
look for solutions of the two equations (\ref{lich}) and 
(\ref{vecteqn}) with $(\phi,W)$ close to $(1,0)$ in appropriate 
weighted H\"older spaces. Set
\[
\X= \left\{(\phi,W,\tau): \phi \in \calC^{k+2,\alpha}_{-\nu}(\RR^n),
\ W \in \calC^{k+1,\alpha}_{-\nu/2}(\RR^n, T\RR^n),\ 
\tau \in \calC^{k,\alpha}_{-\nu/2-1}(\RR^n)\right\}
\]
and
\[
\Y=\calC^{k,\alpha}_{-\nu-2}(\RR^n)\times
\calC^{k-1,\alpha}_{-\nu/2-2}(\RR^n,T\RR^n).
\]
Then it is obvious that
\[
\X \ni (\phi,W,\tau) \longrightarrow \calN(\phi,W,\tau) \in \Y
\]
is a $\calC^1$ mapping in some neighbourhood ${\mathcal U}$ of $(1,0,0) 
\in \X$. Let $\calL$ denote the linearization $D_{12}\calN$ of $\calN$ 
in the $(\phi,W)$ directions, evaluated at the solution $(1,0,0)$. Then
\[
\calL(\psi, Z)=(\Delta \psi, LZ),
\]
and clearly
\begin{equation}
\calL: \calC^{k+2,\alpha}_{-\nu}(\RR^n) \times
\calC^{k+1,\alpha}_{-\nu/2}(\RR^n,T\RR^n)
\longrightarrow \Y.
\label{surjL}
\end{equation}
The fact that this linearization decouples reflects that
we are linearizing about a CMC solution. 

\begin{theorem}
Let $\calN$ denote the vacuum constraint operator as above evaluated at the 
flat metric on $\R^n$ with vanishing extrinsic curvature.
Fix $0<\nu<n-2$. Then for some $\delta > 0$, there exists a $\calC^1$ mapping 
\[
\calF: \{\tau: ||\tau||_{k,\alpha,-\nu/2-1} < \delta\}
\longrightarrow 
\calC^{k+2,\alpha}_{-\nu}(\RR^n)\times
\calC^{k+1,\alpha}_{-\nu/2}(\RR^n,T\RR^n)
\]
such that $\calN(\calF(\tau),\tau) \equiv 0$ (and all solutions near $(1,0,0)$
are of this form). In particular, we may solve the vacuum constraint 
equations for any 
specified mean curvature function $\tau$ with sufficiently small norm 
in $\calC^{k,\alpha}_{-\nu/2-1}(\RR^n)$.
\label{smalltau-thm}
\end{theorem}
\noindent
{\bf Proof:} This follows immediately from the implicit function
theorem once we recognize that for $\nu$ in this range,
$\calL$ is an isomorphism, cf.\ \cite{CCB}, \cite{CB2} for the 
analogous statement on weighted Sobolev spaces. $\hfill\Box$

We shall apply this theorem by choosing a function
$\tau \in \calC^{k,\alpha}_{-\nu/2-1}(\RR^n)$ with
sufficiently small norm, and which is equal to a nonzero
constant on a ball around the origin. Writing
$\calF(\tau) = (\phi,W)$, then $(\phi,W,\tau)$ is
a non-CMC solution of the vacuum constraints.

\begin{proposition} The linearization $\calL = D_{12}\calN$
of the operator $\calN$ in the first two slots, evaluated
at this special solution $(\phi,W,\tau)$, is an isomorphism
 in (\ref{surjL}) when $\delta$ is sufficiently
small. Hence $(\phi,W,\tau)$ is a nondegenerate solution.
\end{proposition}

This follows simply because the invertibility is
an open condition, hence is stable under perturbations of 
small norm.

Finally, notice that we can scale this solution by dilations on 
$\RR^n$ so that $\tau$ is equal to any desired constant on a 
(possibly smaller) ball around the origin. We assume that this has 
been done so that $\tau=n$ on $B_{2R}(0)$ for some $R>0$.  
These dilated  solutions are still nondegenerate. 
We summarize this as
\begin{proposition}
There exist nondegenerate, asymptotically Euclidean, non-CMC solutions of the
vacuum constraint equations $(\gamma, \Pi)$ on $\R^n$ which 
satisfy $\tau\equiv n$ on $B_{2R}(0)\subset \R^n$ for some 
$R>0$. These solutions are given as $(1,W,\tau)$ (or equivalently as
$(1,\mu,\tau)$ with $\mu=\calD W$) relative to the
metric $\phi^{4/(n-2)}\delta$ with $(\phi, W, \tau)\in \X$.
\label{Rn-solns}
\end{proposition}

\section{The approximate solutions}
\label{se:4}
We now sketch the construction of the family of approximate
solutions. This proceeds exactly as in \cite{IMP} and so
we refer to \S2 of that paper for details.

Our two summands are the manifolds $\Sigma$ and $\RR^n$. 
These each have solutions of the vacuum constraints,
which we write as $(\gamma_j, \phi_j, \mu_j, \tau_j)$, $j=1,2$ 
(with $j=1$ corresponding to $\Sigma$ and $j=2$ to 
$\RR^n$). The metric $\gamma_1$ is provided by Proposition~\ref{cmc-compact}
and corresponds to the solution with $\phi_1 \equiv 1$, $\mu_1 \equiv 0$, 
and $\tau_1\equiv n$. Note that if $(\Sigma, \gamma_1)$ has non-trivial 
conformal Killing fields, hence is degenerate, small perturbations
of the conformal class of $\gamma_1$ generically have none.
We then appeal to $\cite{I1}$ to find the corresponding solutions of 
the constraints, which by construction are now nondegenerate.
The solution $(\gamma_2, \phi_2, \mu_2, \tau_2)$ on $\R^n$ is
provided by Proposition~\ref{Rn-solns} with 
$\phi_2 \equiv 1$, $\tau_2\equiv n$ in $B_{2R}(0)\subset\R^n$.
Note that we are including the functions $\phi_j\equiv 1$, for $j=1,2$, 
to emphasize that, relative to the other pieces of data,
these are solutions to the Lichnerowicz equation (\ref{lich})
on each summand.

We begin by removing a small ball of radius $R$
around the point $p \in \Sigma$ and also around the
origin in $\RR^n$. (These points will be written
$p_1$ and $p_2$.) The remaining manifolds 
both have boundaries diffeomorphic to $S^{n-1}$.
The usual connected sum 
construction proceeds by identifying these copies of
$S^{n-1}$, and we denote the resulting manifold
$\widehat{\Sigma} = \Sigma \# \RR^n$. 
The mean curvature functions $\tau_1$ and $\tau_2$ have an obvious
smooth extension, which we denote by $\tau$, 
to all of $\widehat{\Sigma}$.    

We now construct a one-parameter family of metrics $\gamma_T$ and 
symmetric $2$-tensors $\Pi_T$ which serve as background data 
(in the conformal method) for a family of approximate solutions.
More specifically, we construct on  $\widehat{\Sigma}$
a family of metrics $\gamma_T$, functions $\psi_T$ 
(which are equal to $1$ on $(\Sigma\setminus B_R(p))
\cup (\RR^n\setminus B_R(0))$), and trace-free $(0,2)$ tensors  $\mu_T$ 
(which are equal to $0$ on $\Sigma\setminus B_R(p)$ and
$\mu_2=$ on $\RR^n\setminus B_R(0)$, respectively), such that
\[
\calN(\psi_T, 0,\tau) = E_T,
\]
where the error term $E_T$ is exponentially small in $T$
as $T \to \infty$ (see \cite{IMP} for a geometric description of the 
parameter $T$). 
Note that here the constraint operator, $\calN(\cdot,\cdot,\tau)$, is 
computed with respect to $\gamma_T$ and $\mu_T$.
This part of the construction is completely
localized on the `neck region' bridging the two summands. 
These approximate solutions
are perturbed in the next section to a family of exact
solutions of the constraints, and it is only at this last
step that we introduce a global correction term, which is
exponentially small in $T$. 

To construct these approximate solutions we first 
choose conformal factors $\psi_j$ on each of the summands
which are identically one outside the balls $B_{3R/2}(p_j)$ and which
are equal to $(\mbox{dist}\,(\cdot, p_j))^{(n-2)/2}$ in $B_{R}(p_j)$. 
We then define
\[
(\g_j)_c= \psi_j^{-\frac{4}{n-2}}\g_j.
\]
These are complete metrics with asymptotically cylindrical
ends in place of the punctured balls $B_R(p_j)$. 
There is a decomposition of the $\Pi_j$ into trace-free and pure-trace 
parts,  
\[
\Pi_1 = \frac{\tau_1}{n} \g_1, \qquad
\Pi_2 = \mu_2 + \frac{\tau_2}{n} \g_2,
\]
(Notice that $\mu_2$ is not transverse-traceless: it has non-vanishing 
divergence since $\nabla \tau_2\not\equiv 0$.) 
Let $(\mu_2)_c=\psi_2^2 \mu_2$ and set $(\mu_1)_c=\mu_1=0$.
The discussion from \S 2 shows that $((\g_j)_c, \psi_j, (\mu_j)_c, \tau_j)$
are solutions (with complete metrics) of the constraint equations  
on $\Sigma\setminus \{p\}$ and $\RR^n \setminus \{0\}$, respectively. 

If $r_j = \mbox{dist}_{\gamma_j}(\cdot, p_j)$, then $t_j = -\log r_j$
is a natural linear coordinate on each of these cylindrical ends.
Let $T$ be a large parameter and $A=-\log R$. Truncate the 
cylindrical ends of each of these manifolds by omitting the regions
where $t_j > A + T$. A smooth manifold diffeomorphic to $\widehat{\Sigma}$
is obtained by identifying the two finite cylindrical segments 
$\{(t_j,\theta): A \leq t_j \leq A+T\}$ via the map $(t_1,\theta) 
\to (T-t_1,-\theta)$. We call this new manifold $\ST$, and denote
the long cylindrical tube it contains by $C_T$. It is convenient
to use 
\[
s = t_1 - A - T/2 = -t_2 + A + T/2.
\]
as a linear coordinate on $C_T$. This parametrizes $C_T$ via
the chart $(s,\theta) \in [-T/2,T/2] \times S^{n-1}$.

We next define the family of metrics $\gt$ and trace-free tensors $\mut$
on $\ST$. Choose cutoff functions $\chi_1,\chi_2$ so that
$\chi_1 = 1$ on all of $\Sigma$ and vanishes in the ball
$B_{e^{-T/2}R}(p_1)$ and similarly for $\chi_2$, and moreover
such that when these functions are transferred to $\ST$,
then $\chi_2 = 1-\chi_1$ and the supports of $d\chi_j$ are
contained in the region $Q = [-1,1]_s \times S^{n-1}_\theta
\subset C_T$. Now define
\[
\gamma_T = \chi_1 (\gamma_1)_c + \chi_2 (\gamma_2)_c, \qquad
\mu_T = \chi_2 (\mu_2)_c.
\]
Notice that $\gamma_T = (\gamma_j)_c$ and $\mu_T = (\mu_j)_c$,
where $j=1$ when $s \leq -1$ and on the rest of $\Sigma$ and
$j=2$ when $s \geq 1$ and on the rest of $\RR^3$. 
The key point to emphasize here is that
$(\mu_2)_c$ is very close to zero in the region where we 
have cut it off to be exactly zero, so this introduces only 
a small error.
This is described in more detail in \S2 of \cite{IMP}.

The conformal factors $\psi_j$ on either summand must be
joined together somewhat differently, by cutting them
off at the far ends of the cylinder (relative to their
domain of definition), as follows.  We choose
nonnegative cutoff functions $\tilde{\chi}_1$ on $\Sigma \setminus 
\{p\}$ and $\tilde{\chi}_2$ on $\RR^n \setminus \{0\}$ which are
identically one for $t_j \leq A+T-1$ and which vanish
when $t_j \geq A+T$. Now set
\[
\psi_T = \tilde{\chi}_1 (\psi_1)_c + \tilde{\chi}_2 (\psi_2)_c.
\]
This is defined on $\ST$, is identically $1$ away from $C_T$, and  
equals $\psi_1+\psi_2$ on most of the cylinder except at the ends.

The estimates for the error term $E_T = \calN(\psi_T, 0, \tau)$
now follow readily from the estimates in \S 3.4, \S 4 and \S 6 of \cite{IMP}.
In particular, writing $E_T =(E^1_T, E^2_T)\in\Y$ (corresponding
to (\ref{lich}) and (\ref{vecteqn}), respectively) we have that 
\begin{equation}
\|E^1_T\|_{k,\alpha,-\nu-2} + \|E^2_T\|_{k-1,\alpha,-\nu/2-2} 
\leq C e^{-T/4},
\label{ests}
\end{equation}
and furthermore, the components $E^1_T$ and $E^2_T$ of $E_T$ are 
supported on all of $C_T$ and $Q \subset C_T$, respectively.  
In establishing this estimate it is important to note that, since 
$\nabla\tau\equiv 0$ on $C_T$, the vector constraint
equation (\ref{vecteqn}) does not introduce any new error terms beyond 
those previously encountered in \cite{IMP}.

\section{The perturbation result}
\label{se:5}

We now sketch the argument to perturb the approximate
solution 
\[
\widetilde\gamma_T=\psi_T^{\frac{4}{n-2}}\gamma_T,\qquad 
\widetilde\Pi_T=\psi_T^{-2}\mu_T +
\frac{\tau}{n}\psi_T^{\frac{4}{n-2}}\gamma_T
\] 
on $\Sigma_T$, to an exact, asymptotically
Euclidean solution $(\bar\gamma_T,\bar\Pi_T)$ of the vacuum constraints,
corresponding to a solution $(\phi_T, Z_T)$ to $\calN(\phi_T,Z_T,\tau)=0$
on $\ST$ (where $\calN(\cdot,\cdot,\tau)$ is computed relative to
$\gamma_T$ and $\mu_T$), when $T$ is sufficiently large. Here
$\phi_T=\psi_T +\phi$ and the pair $(\phi, Z_T)$ are small in 
the appropriate function space.

\begin{definition}
Let $w_T$ be an everywhere positive smooth function on $\Sigma_T$ 
which equals $e^{-T/4}\cosh(s/2)$ on $C_T$ and which equals
$1$ outside both balls $B_{2R}(p_j)$. For any $\delta\in \R$, and any $\phi
\in {\mathcal C}^{k,\alpha}_{\nu'}(\ST)$, set 
\[
||\phi||_{k,\alpha,\nu',\delta} = ||w_T^{-\delta}\phi||_{k,\alpha,\nu'};
\]
the corresponding space is denoted ${\calC}^{k,\alpha}_{\nu',\delta}
(\ST)$. \label{ct-weight}
\end{definition}

Thus elements of this function space not only have restricted
growth or decay at infinity, but their norms are also measured
in $C_T$ with an extra weighting factor. We let
\[
\X_{T,\delta} = 
\left\{(\phi,Z): \phi \in \calC^{k+2,\alpha}_{-\nu,\delta}(\ST),
\ Z \in \calC^{k+1,\alpha}_{-\nu/2}(\ST; T\ST)
\right\}
\]
and
\[
\Y_{T,\delta}=\calC^{k,\alpha}_{-\nu-2,\delta}(\ST)\times
\calC^{k-1,\alpha}_{-\nu/2-2}(\ST; T\ST).
\]
Notice that we are only including the weight $\delta$ on the first 
component of this space, but do not measure the vector field $Z$ 
with a weighted norm in the neck.
We use the obvious product norm on $\X_{T,\delta}$, but
the less obvious one
\[
||(f,Y)||_{\Y_{T,\delta}} =
||f||_{k,\alpha,-\nu-2,\delta} + T^{-3}||Y||_{k-1,\alpha,-\nu/2-2}
\]
on $\Y_{T,\delta}$.

We also assume that the weight $\nu$ is always in the range $(0,n-2)$,
so that the conclusion of Proposition 2 is valid. 

The mapping
\[
\X_{T,\delta} \ni (\phi,Z) \longrightarrow
\tilde{\calN}(\phi,Z) \equiv 
\calN(\psi_T + \phi, Z,\tau) \in \Y_{T,\delta}
\]
is $\calC^1$ in some small neighbourhood ${\mathcal U}$
of the origin in $\X_{T,\delta}$ for each $T$. The only subtlety
here is that even when it has small norm, the function $\phi$ 
may be rather large in a pointwise sense in $C_T$. However,
$\nabla \tau = 0$ in $C_T$ and so this does not affect
(\ref{vecteqn}) there. Furthermore, if we write
the linearization of $\tilde{\calN}$ at $(\phi,Z) = (0,0)$ as 
$\calL_T$, then 
\[
\calL_T: \X_{T,\delta}  \longrightarrow \Y_{T,\delta}
\]
is bounded as well.

We state the two fundamental results, which are essentially obtained 
by combining Propositions 7 and 8 from \S 5 and Corollary 1 from
\S 3.3 in \cite{IMP}.

\begin{proposition}
Fix any $\delta\in\R$. For $T$ sufficiently large, the mapping 
\[
\calL_T : \X_{T,\delta} \longrightarrow  \Y_{T,\delta}
\]
is an isomorphism.
\label{pr:2}
\end{proposition}
Let $\calG_T$ denote the inverse of $\calL_T$ provided by
this proposition. Thus
\begin{equation}
\calG_T: \Y_{T,\delta}\longrightarrow \X_{T,\delta}
\label{eq:mapgt}
\end{equation}
and $\calL_T \calG_T = \calG_T \calL_T = I$. 
Of course $\calG_T$ also depends on $\delta$, but we suppress this 
in the notation. 

\begin{proposition}
If $0 < \delta < 1$, then the norm of 
${\calG}_T$ is uniformly bounded as $T \to \infty$.
\label{pr:3}
\end{proposition}

Although we refer to \cite{IMP} for the proofs, let
us make a few comments. Proposition~\ref{pr:2} reflects
the fact that the solutions we are joining are nondegenerate
on their respective summands. One way to prove this result is
to patch together the (pseudodifferential) inverses on
each piece with the inverse on $C_T$ (which is constructed rather 
explicitly in \cite{IMP}). The resulting parametrix $G_T$ satisfies
$\calL_T G_T = I - R_T$, where $R_T$ has very small norm.
As for Proposition~\ref{pr:3}, the reason we have added the
$T^{-3}$ factor in the second component in the definition
of $Y_{T,\delta}$ is because the inverse of the vector
Laplacian $L = \calD^* \calD$ localized to the neck region $C_T$
has norm bounded by $T^3$ (cf. \S 3.3 of \cite{IMP}). 
This proposition is most
handily proved by assuming this uniform bound fails and 
arguing to a contradiction.  

From here, the rest of the proof of Theorem~1 is straightforward.
The system we wish to solve is written as 
\[
\calN(\psi_T + \phi, Z,\tau) =0.
\]
We write this as 
\[
\calL(\phi, Z) = F_T(\phi, Z)
\]
where $F_T$ depends on all of the approximate
data and consists of the error term $E_T$ together with a nonlinear
operator which is quadratically small. Using Proposition~\ref{pr:2}
this may in turn be written as 
\[
(\phi, Z) =\calG_T(F_T(\phi, Z))
\]
The existence of a fixed point for this map then follows from an
application of the contraction mapping principle using 
Proposition~\ref{pr:3} and the error estimate (\ref{ests}). This
is explained carefully in \S 6 of \cite{IMP}. This completes
the proof of Theorem 1.

We emphasize that the reason no substantial changes need to be 
made to any of these arguments is that their most difficult
aspects involve the explicit analysis of the linearizations
of the Lichnerowicz operator (\ref{lich}) and the vector Laplacian 
(\ref{vecteqn}) in $C_T$ and the estimates of the approximate
solutions in this same region. Because we are always assuming
that $\nabla \tau = 0$ there, these parts of the arguments
carry through completely unchanged. The remaining more global
parts are quite general and it can be readily verified that
they do not `notice' the fact that these operators are now
coupled. 

Very few modifications are required to prove the more general 
Theorem 2. In fact, we need only note that now each of the 
initial solutions may have a transverse-traceless part,
but these are easily incorporated into all the arguments. 

\section{Asymptotically flat vacuum spacetimes with no maximal slices}
\label{se:6}

It has been commonly believed that every physically reasonable asymptotically 
flat solution of the Einstein equations should admit a foliation by 
asymptotically Euclidean, maximal ($\tau=0$) slices. Such foliations are 
unique when they exist, and they are useful because they have desirable 
``singularity avoidance'' properties. 
Roughly 20 years ago, Brill showed \cite{Br} that there exist 
asymptotically flat spacetimes which are {\it dust} solutions of the 
Einstein equations and which admit no maximal spacelike hypersurfaces. 
These are constructed by explicitly gluing a Friedman-Robertson-Walker 
spacetime to a Schwarzschild spacetime. The
presence of dust is crucial to this argument.

Our analysis here shows that there exist asymptotically flat
{\em vacuum} spacetimes which admit 
no maximal Cauchy slices (satisfying certain decay conditions).
Indeed, if $(M,g)$ is the maximal development of any initial data
set constructed as in Theorem~1, with $\Sigma$ admitting no metric of 
positive 
scalar curvature, then $(M,g)$ admits no such maximal slice.
As mentioned earlier, such manifolds $\Sigma$ are quite abundant; 
for example, any closed hyperbolic 3-manifold has this property.  
The precise result is as follows.  

\begin{theorem}
Suppose that $\Sigma$ is a closed $n$-manifold which admits no metric of 
positive scalar curature. For any $p\in \Sigma$, let $(\gamma,\Pi)$ be  
an asymptotically Euclidean solution of the vacuum constraint equations 
on $\Sigma\setminus\{p\}$ provided by Theorem~1, and let $(M,g)$ be the 
maximal development of this data. Then there exists no maximal ($\tau=0$) 
asymptotically Euclidean Cauchy surface $\widehat\Sigma$ in $M$ 
for which the induced metric $\hat\gamma \in \calM^{k,\alpha}_{-\nu}$ 
for some $\nu> \frac{n-2}{2}$, and for which the scalar curvature 
$R(\hat\gamma)\in L^1(\widehat\Sigma)$. 
\label{thm:maxslice}
\end{theorem}
\noindent
{\bf Proof:}
We have already observed that any maximal slice 
$(\widehat{\Sigma},\hat{\gamma})$ has nonnegative scalar curvature. 
Schoen and Yau have proved, cf.\ \cite{LP} or 
\cite{S}, that an asymptotically Euclidean manifold with nonnegative 
scalar curvature may be perturbed to an asymptotically Euclidean 
manifold which is scalar flat and in addition conformally flat near 
infinity. Let $\gamma'$ denote this new metric. Then there exists a 
compact set $K\subset \widehat\Sigma$ which is diffeomorphic to an 
exterior region in $\RR^n$ such that in the associated
Euclidean coordinates, $\gamma'=u(x)^{\frac{4}{n-2}}\delta$ 
with $u(x)\rightarrow 1$ as $|x|\rightarrow\infty$ ($\delta$ is 
the Euclidean metric).  Since both $\delta$ and $\gamma'$ are 
scalar flat, $u(x)$ is harmonic and thus 
has an expansion
\[
u(x)=1 + E\, |x|^{2-n} +\mathcal{O} (|x|)^{1-n}, \qquad |x|\to \infty;
\]
here $E$ is the total energy of $\gamma'$, see \cite{S}. The Kelvin 
transform $\bar{u}(x)= |x|^{2-n}u(|x|^{-2} x)$ of $u$ is harmonic on  some
punctured ball $B(0,\rho)\setminus\{0\}$. Hence $(\widehat\Sigma, \gamma')$ 
may be conformally compactified as follows. Choose a positive smooth 
function $G$ on $\widehat\Sigma$ so that $G(x)^{-1}= \bar u(x)$ when 
$|x|> 2R$ and $G\equiv 1$ on $K$.
Then $\bar\gamma= G^{-\frac{4}{n-2}}\gamma'$ extends smoothly to 
$\Sigma\cong \widehat\Sigma\cup\{\infty\}$. Regard $G$ as a function on
$\Sigma \setminus \{p\}$. The facts that it is positive,  
diverges near $p$ like
$\mbox{dist}(\cdot, p)^{2-n}$, and is in the nullspace of the
conformal Laplacian away from $p$ (since $\gamma'$ is scalar flat)
shows that it is a constant multiple of Green's function for the 
conformal Laplacian 
(for $\overline{\gamma}$) on $\Sigma$. The positivity of Green's function
is equivalent to the positivity of the first eigenvalue of the conformal 
Laplacian for $\overline{\gamma}$, which in turn implies that this metric 
is conformally equivalent to a metric on $\Sigma$ with positive 
scalar curvature. This is a contradiction and so the proof is finished. 
$\hfill\Box$

\begin{remark} This equivalence between asymptotically Euclidean metrics of 
nonnegative scalar curvature and metrics of positive scalar curvature on 
`stereographic compactifications' is well-known, although the 
proof does not seem to be readily available in the literature.  
Note that by choosing mean curvature functions $\tau$ on $\R^n$ with
sufficiently fast decay one can use Theorem~1 to produce initial 
data sets which
satisfy the asymptotic conditions required in Theorem \ref{thm:maxslice}.
It is not at all clear, however, and may be quite subtle to prove, 
that any other asymptotically Euclidean Cauchy surface in the resulting 
maximal development must also satisfy these same decay conditions.
\end{remark}


\begin{thebibliography}{999}

\bibitem{AC} L. Andersson and P. T.\ Chru\'sciel,
{\em Solutions of the constraint equations in general relativity 
satisfying
``hyperboloidal boundary conditions'',}
Dissertationes Math. (Rozprawy Mat.) {\bf 355} (1996)

\bibitem{A} T. Aubin,
{\em M\' etricques Riemanniennes et courbure,}
J. Differential Geometry {\bf 4} (1970) 383--424.

\bibitem{Bar} R. Bartnik,
{\em The mass of an asymptotically flat manifold,}
Comm. Pure Appl. Math. {\bf 39} (1986) 661--693.

\bibitem{Br} D. Brill,
{\em On Spacetimes without Maximal Surfaces,}
in ``Proceedings of the Third Marcel Grossman Meeting'', 
Hu Ning, editor, Science Press and 
North Holland Publ. Co. (1983)  79--87.


\bibitem{Ca1} M. Cantor,
{\em The existence of non-trivial asymptotically flat initial
data for vacuum spacetimes,}
Comm. Math. Phys. {\bf 57} (1977), no. 1, 83--96.

\bibitem{CB} Y. Choquet-Bruhat,
{\em Th\' eor\` eme d'existence pour certains syst\` emes d'\' equations
aux d\' eriv\' ees partialles non lin\' eaires,}
Acta. Math. {\bf 88} (1952) 141--225.

\bibitem{CB2} Y. Choquet-Bruhat,
{\em Solution of the coupled Einstein constraints on asymptotically
Euclidean manifolds,} in ``Directions in General Relativity'', Vol. 2,
B.~L.~Hu and T.~A.~Jacobson ed. Cambridge University Press, Cambridge 
(1993)
83--96.

\bibitem{CIY} Y. Choquet-Bruhat, J. Isenberg and J. York,
{\em Einstein constraints on asymptotically Euclidean manifolds,}
Phys. Rev. D (3) {\bf 61} (2000), no. 8, 084034, 20 pp.


\bibitem{CCB} D. Christodoulou and Y. Choquet-Bruhat,
{\em Elliptic systems in $H_{s,\delta}$ spaces on manifolds which are
Euclidean at infinity},
Acta. Math. {\bf 146} (1981) 129--150.

\bibitem{COM} D. Christodoulou and N. O'Murchadha,
{\em The boost problem in general relativity,}
Comm. Math. Phys. {\bf 80} (1981) 271--300.

\bibitem{Cor} J. Corvino,
{\em Scalar curvature deformation and a gluing construction
for the Einstein constraint equations,}
Comm. Math. Phys. {\bf 214} (2000), no. 1, 137--189.

\bibitem{GL1} M. Gromov and H.B. Lawson,
{\em Spin and scalar curvature in the presence of a fundamental group I,}
Ann. of Math. {\bf 111}  (1980), no. 2, 202--230.

\bibitem{GL2} M. Gromov and H.B. Lawson,
{\em The classification of simply connected manifolds of positive scalar
curvature,}
Ann. of Math. {\bf 111}  (1980), no. 3, 423--434.

\bibitem{I1} J. Isenberg,
{\em Constant mean curvature solutions of the Einstein constraint
equations on closed manifolds,}
Class. Quantum Grav. {\bf 12} (1995) 2249--2274.

\bibitem{IMaxP} J. Isenberg, D. Maxwell and D. Pollack, In preparation.

\bibitem{IMP} J. Isenberg, R. Mazzeo and D. Pollack,
{\em Gluing and Wormholes for the Einstein constraint equations,}
gr-qc/0109045, To appear Comm. Math. Phys.


\bibitem{IM} J. Isenberg and V. Moncrief,
{\em A set of nonconstant mean curvature solutions of the Einstein 
constraint
equations on closed manifolds,}.
Class. Quantum Grav. {\bf 13} (1996) 1819--1847.

\bibitem{KW} J. Kazdan and F. Warner,
{\em Scalar curvature an conformal deformation of Riemannian structure,}
J. Differential Geom. {\bf 10} (1975) 113--134.

\bibitem{LP} J. Lee and T. Parker,
{\em The Yamabe probnlem,}
Bull. Amer. Math. Soc. {\bf 17} (1987) 37--81.

\bibitem{RS} J. Rosenberg and S. Stolz,
{\em Manifolds of positive scalar curvature,} in
``Algebraic Topology and its Applications'' {\bf 27} (1994)
G.E. Carlsson et al., eds.,  Math. Sci. Res. Inst. Publ.
Springer-Verlag, NY, 241--267.

\bibitem{S} R. Schoen,
{\em Variational theory for the total scalar curvature functional 
for Riemannian metrics and related topics,}
``Topics in Calculus of Variations'' LNM 1365, Springer-Verlag,  
M. Giaquinta ed. (1987) 120--154.

\bibitem{SY1} R. Schoen and S.T. Yau,
{\em On the structure of manifolds with positive scalar curvature,}
Manuscripta Math. {\bf 28} (1979) 159--183.

\bibitem{SY2} R. Schoen and S.T. Yau,
{\em On the proof of the positive mass conjecture in general relativity,}
Comm. Math. Phys. {\bf 65} (1979) 45--76.

\bibitem{Wi} D. Witt, {\em Vacuum space-times that admit no maximal 
slice,}
Phys. Rev. Let. {\bf 75} No. 12 (1986) 1386--1389.

\end{thebibliography}
\end{document}